\catcode`\@=11 
\def\lsim{\mathrel{\mathpalette\@versim<}}
\def\gsim{\mathrel{\mathpalette\@versim>}}
\def\@versim#1#2{\vcenter{\offinterlineskip
        \ialign{$\m@th#1\hfil##\hfil$\crcr#2\crcr\sim\crcr } }}

\documentstyle[preprint,aps]{revtex}
\begin{document}

\newcommand{\ee}{\mbox{$e^+e^-$}}
\newcommand{\phinaught}{\mbox{$\langle\Phi\rangle_0$}}
\newcommand{\mt}{m_t}
\newcommand{\mh}{m_H}
\newcommand{\mgev}{GeV/c$^2$}
\newcommand{\mevocsq}{MeV/c$^2$}
\newcommand{\mum}{$\mu$m}
\newcommand{\roots}{\sqrt{s}}
\newcommand{\alr}{A_{LR}}
\newcommand{\alro}{A_{LR}^0}
\newcommand{\swein}{\sin^2\theta_W^{\rm eff}}
\newcommand{\pole}{{\cal P}_e}
\newcommand{\polp}{{\cal P}_p}
\newcommand{\pollin}{{\cal P}_e^{lin}}
\newcommand{\pollum}{\bar{\cal P}_e}
\newcommand{\polg}{{\cal P}_\gamma}
\newcommand{\poll}{{\cal P}}
\newcommand{\apolel}{\langle{\cal P}_e\rangle}
\newcommand{\ameas}{A_m}
\newcommand{\lum}{{\cal L}}
\newcommand{\alum}{A_{\cal L}}
\newcommand{\aeff}{A_\varepsilon}
\newcommand{\apol}{A_{\cal P}}
\newcommand{\apower}{{\cal A}}
\newcommand{\aengy}{A_E}
\newcommand{\aback}{A_b}
\newcommand{\etal}{{\it et al.}}
\newcommand{\zo}{Z}
\newcommand{\rar}{\rightarrow}

\draft

\preprint{\vbox{\hsize=120pt\noindent SLAC--PUB--7291 \\
November 1996 }}

\title{An Improved Measurement of the Left-Right Z$^0$ Cross Section
Asymmetry$^\dagger$}
\author{The SLD Collaboration$^*$}
\address{Stanford Linear Accelerator Center\\
         Stanford University, Stanford, California, 94309\\}
\maketitle

\begin{abstract}
We present a new measurement of
the left-right cross section asymmetry ($\alr$)
for $\zo$ boson production
by \ee\ collisions.  The measurement was performed
at a center-of-mass energy of 91.28~GeV
with the SLD detector at the SLAC Linear Collider (SLC).
The luminosity-weighted average polarization of the SLC
electron beam was (77.23$\pm$0.52)\%.
Using a sample of 93,644 $\zo$ decays, we measure the pole-value
of the asymmetry, $\alro$, to be
0.1512$\pm$0.0042({\rm stat.})$\pm$0.0011({\rm syst.}) which is equivalent
to an effective weak mixing angle of
$\swein=0.23100\pm0.00054({\rm stat.})\pm0.00014({\rm syst.})$.

\end{abstract}
\vskip 0.3in
\begin{center}
{\rm Submitted to {\em Physical Review Letters}}
\end{center}
\vskip 1.3in
\vbox{\footnotesize\renewcommand{\baselinestretch}{1}\noindent
 $^\dagger$This work was supported in part by Department of Energy
  contract DE-AC03-76SF00515}
\pagebreak
\begin{center}
\bigskip
%
%
%
  \def\iADEL{$^{(1)}$}
  \def\iBOL{$^{(2)}$}
  \def\iBU{$^{(3)}$}
  \def\iBRUN{$^{(4)}$}
  \def\iUCSB{$^{(5)}$}
  \def\iUCSC{$^{(6)}$}
  \def\iCIN{$^{(7)}$}
  \def\iCSU{$^{(8)}$}
  \def\iCOLO{$^{(9)}$}
  \def\iCOL{$^{(10)}$}
  \def\iFER{$^{(11)}$}
  \def\iFRA{$^{(12)}$}
  \def\iILL{$^{(13)}$}
  \def\iLBL{$^{(14)}$}
  \def\iMIT{$^{(15)}$}
  \def\iMASS{$^{(16)}$}
  \def\iMISS{$^{(17)}$}
  \def\iMOSC{$^{(18)}$}
  \def\iNAG{$^{(19)}$}
  \def\iOREG{$^{(20)}$}
  \def\iPAD{$^{(21)}$}
  \def\iPERU{$^{(22)}$}
  \def\iPISA{$^{(23)}$}
  \def\iRUT{$^{(24)}$}
  \def\iRAL{$^{(25)}$}
  \def\iSOGANG{$^{(26)}$}
  \def\iSOONG{$^{(27)}$}
  \def\iSLAC{$^{(28)}$}
  \def\iTENN{$^{(29)}$}
  \def\iTOH{$^{(30)}$}
  \def\iVAND{$^{(31)}$}
  \def\iWASH{$^{(32)}$}
  \def\iWISC{$^{(33)}$}
  \def\iYALE{$^{(34)}$}
  \def\dead{$^{\dag}$}
  \def\andgen{$^{(a)}$}
  \def\andper{$^{(b)}$}
%
%
$^*$
\mbox{K. Abe                 \unskip,\iNAG}
\mbox{K. Abe                 \unskip,\iTOH}
\mbox{I. Abt                 \unskip,\iILL}
\mbox{T. Akagi               \unskip,\iSLAC}
\mbox{N.J. Allen             \unskip,\iBRUN}
\mbox{W.W. Ash               \unskip,\iSLAC$^\dagger$}
\mbox{D. Aston               \unskip,\iSLAC}
\mbox{K.G. Baird             \unskip,\iMASS}
\mbox{C. Baltay              \unskip,\iYALE}
\mbox{H.R. Band              \unskip,\iWISC
% ...
}
\it
  \vskip \baselineskip                   
  \vskip \baselineskip                   
%
%
%
  \iADEL
     Adelphi University,
     Garden City, New York 11530 \break
  \iBOL
     INFN Sezione di Bologna,
     I-40126 Bologna, Italy \break
  \iBU
     Boston University,
     Boston, Massachusetts 02215 \break
  \iBRUN
     Brunel University,
     Uxbridge, Middlesex UB8 3PH, United Kingdom \break
  \iUCSB
     University of California at Santa Barbara,
     Santa Barbara, California 93106 \break
  \iUCSC
     University of California at Santa Cruz,
     Santa Cruz, California 95064 \break
  \iCIN
     University of Cincinnati,
     Cincinnati, Ohio 45221 \break
  \iCSU
     Colorado State University,
     Fort Collins, Colorado 80523 \break
  \iCOLO
     University of Colorado,
     Boulder, Colorado 80309 \break
  \iCOL
     Columbia University,
     New York, New York 10027 \break
  \iFER
     INFN Sezione di Ferrara and Universit\`a di Ferrara,
     I-44100 Ferrara, Italy \break
  \iFRA
     INFN  Lab. Nazionali di Frascati,
     I-00044 Frascati, Italy \break
  \iILL
     University of Illinois,
     Urbana, Illinois 61801 \break
  \iLBL
     Lawrence Berkeley Laboratory, University of California,
     Berkeley, California 94720 \break
  \iMIT
     Massachusetts Institute of Technology,
     Cambridge, Massachusetts 02139 \break
  \iMASS
     University of Massachusetts,
     Amherst, Massachusetts 01003 \break
  \iMISS
     University of Mississippi,
     University, Mississippi  38677 \break
  \iMOSC
    Moscow State University,
    Institute of Nuclear Physics
    119899 Moscow, Russia    \break
  \iNAG
     Nagoya University,
     Chikusa-ku, Nagoya 464 Japan  \break
  \iOREG
     University of Oregon,
     Eugene, Oregon 97403 \break
  \iPAD
     INFN Sezione di Padova and Universit\`a di Padova,
     I-35100 Padova, Italy \break
  \iPERU
     INFN Sezione di Perugia and Universit\`a di Perugia,
     I-06100 Perugia, Italy \break
  \iPISA
     INFN Sezione di Pisa and Universit\`a di Pisa,
     I-56100 Pisa, Italy \break
  \iRUT
     Rutgers University,
     Piscataway, New Jersey 08855 \break
  \iRAL
     Rutherford Appleton Laboratory,
     Chilton, Didcot, Oxon OX11 0QX United Kingdom \break
  \iSOGANG
     Sogang University,
     Seoul, Korea \break
  \iSOONG
     Soongsil University,
     Seoul, Korea  156-743 \break
  \iSLAC
     Stanford Linear Accelerator Center, Stanford University,
     Stanford, California 94309 \break
  \iTENN
     University of Tennessee,
     Knoxville, Tennessee 37996 \break
  \iTOH
     Tohoku University,
     Sendai 980 Japan \break
  \iVAND
     Vanderbilt University,
     Nashville, Tennessee 37235 \break
  \iWASH
     University of Washington,
     Seattle, Washington 98195 \break
  \iWISC
     University of Wisconsin,
     Madison, Wisconsin 53706 \break
  \iYALE
     Yale University,
     New Haven, Connecticut 06511 \break
  \dead
     Deceased \break
  \andgen
     Also at the Universit\`a di Genova \break
  \andper
     Also at the Universit\`a di Perugia \break
\rm
%

\end{center}

In 1993, the SLD Collaboration performed a precise measurement
of the left-right cross section asymmetry in the
production of $\zo$ bosons by \ee\ collisions \cite{alr93}.  In this letter,
we present a substantially improved measurement based upon new data
recorded during the 1994/95 run of the SLAC Linear Collider (SLC) with larger
beam polarization and better control of systematic uncertainties.

The left-right asymmetry is defined as
$\alro\equiv\left(\sigma_L-\sigma_R\right)/
\left(\sigma_L+\sigma_R\right)$,
where $\sigma_L$ and $\sigma_R$ are the $\ee$ production
cross sections for $\zo$ bosons at the $\zo$-pole energy
with left-handed and right-handed
electrons, respectively.  The Standard Model predicts
that this quantity depends upon the effective vector ($v_e$)
and axial-vector ($a_e$) couplings of the $\zo$ boson to the electron
current,
\begin{equation}
\alro=\frac{2v_ea_e}{v_e^2+a_e^2}\equiv \frac{
2\left[1-4\swein\right]}{1+\left[1-4\swein\right]^2}, \label{eq:alrswein}
\end{equation}
where the effective electroweak mixing parameter
is defined \cite{swdef}
as $\swein\equiv(1-v_e/a_e)/4$.
Note that $\alro$ is a sensitive function of $\swein$
and depends upon virtual
electroweak radiative corrections including those which involve
the top quark and Higgs boson and those arising from new phenomena.
The recent measurement of the top quark mass \cite{top} has, as a
determination of a previously unknown parameter of the Standard
Model, greatly enhanced the power of this measurement as a test
of the prevailing theory.

We measure the left-right asymmetry by counting hadronic and (with low efficiency) $\tau^+\tau^-$ final states produced in \ee\ collisions
near the $\zo$-pole energy for each of the two longitudinal polarization
states of the electron beam.  The asymmetry formed from these rates,
$\alr$, must then be corrected for residual effects arising from pure photon
exchange and $\zo$-photon interference to extract $\alro$.  The measurement requires
knowledge of the absolute beam polarization, but
does not require knowledge of the
absolute luminosity,
detector acceptance, or efficiency \cite{accept}.

The operation of the SLC with a polarized electron beam
has been described previously \cite{oldslc}.  In 1994,
the beam polarization at the SLC source \cite{source} was increased from 63\%
to $\sim 80\%$
by the use of a thinner (0.1 $\mu$m)
strained-lattice GaAs photocathode \cite{strlat}
which was illuminated by a pulsed
Ti:Sapphire laser operating at 845~nm.
The circular polarization state of each
laser pulse (and hence, the helicity of each electron pulse)
was chosen randomly.
The electron spin orientation
was manipulated in the SLC North Arc by a pair of large
amplitude betatron oscillations to
achieve longitudinal polarization at the SLC interaction point (IP)
\cite{rot}.
The maximum luminosity of the collider was approximately
6$\times$10$^{29}$~cm$^{-2}$sec$^{-1}$.
The luminosity-weighted mean $\ee$ center-of-mass energy ($E_{cm}$) is
measured with
precision energy spectrometers \cite{enspa}
to be 91.280$\pm$0.025~GeV.

The longitudinal electron beam polarization ($\pole$)
is measured by a Compton scattering
polarimeter \cite{polarimeter}
located 33~m downstream of the IP.
After it passes through the IP
and before it is deflected by dipole magnets,
the electron beam collides with a circularly polarized photon
beam produced by a pulsed frequency-doubled Nd:YAG laser of wavelength
532~nm operating at $\sim$17~Hz.  Since the accelerator produces electron pulses at 120~Hz, the polarimeter samples each seventh machine pulse.  The
scattered and unscattered components of the electron
beam remain unseparated until they pass
through a dipole-quadrupole spectrometer.  The
scattered electrons
are dispersed horizontally and exit the vacuum system through
a thin window.  A multichannel
Cherenkov detector observes the scattered
electrons in the interval from 17 to 30~GeV/c.

The counting rates in each detector channel are
measured for three combinations of electron and photon beam parameters: parallel electron and photon helicities, antiparallel helicities, and photon beam absent.  The latter combination is used to measure detector background.
The asymmetry formed from the background-subtracted counting rates is equal to the product
$\pole\polg\apower_i$ where $\polg$ is the circular polarization of the
laser beam at the electron-photon crossing point
and $\apower_i$ is the analyzing power of the $i^{th}$ detector channel.  The laser polarization was maintained at
(99.6$\pm$0.2)\% by continuously monitoring and correcting phase shifts
in the laser transport system.
The analyzing powers of the detector channels incorporate resolution and spectrometer effects and differ slightly from the theoretical Compton
asymmetry function at the mean accepted energy for each channel \cite{comref}.
The minimum energy of a Compton-scattered electron for the initial electron and photon energies is 17.36~GeV.  The location of this kinematic endpoint at the detector was monitored by frequent scans of the detector horizontal position during polarimeter operation.  This technique determines and monitors the analyzing powers of each detector channel.

Polarimeter data are acquired continually during the operation
of the SLC.  The absolute statistical precision attained in a 3 minute measurement is typically $\delta\pole=0.8\%$.
The systematic uncertainties that affect the polarization measurement
are summarized in Table~\ref{table1}.
The total relative systematic uncertainty
is estimated to be
$\delta\pole/\pole=0.64\%$.

Due to energy-spread-induced spin diffusion in the SLC arc
and imperfect spin orientation, the longitudinal polarization
of the electron beam at the IP was typically 98\% of the polarization
in the linac.  This estimate follows from a measurement of the arc
spin rotation matrix performed with a beam of very small energy spread
($\lsim0.05\%$) using a pair of
spin rotation solenoids and the Compton polarimeter.
The electron polarization in the linac was determined
to be (78.6$\pm$0.9)\% and was consistent with a
direct measurement using a diagnostic
M\o ller polarimeter \cite{levchuk} of (81$\pm$3)\%.

In our previous Letter \cite{alr93},
we examined an effect that causes the beam polarization
measured by the Compton Polarimeter, $\pole$, to
differ from the luminosity-weighted beam polarization,
$\pole(1+\xi)$, at the SLC IP.
While the Compton polarimeter measures the polarization of the entire electron bunch, chromatic aberrations in the SLC final focus optics reduce the contribution of off-energy electrons to the luminosity.  The on-energy electrons with larger average longitudinal polarization
therefore
contribute more to the total luminosity and $\xi$ can be non-negligible.
To first order, the magnitude of $\xi$ depends quadratically on
the width of the beam energy distribution $N(E)$,
the energy dependence of the arc spin rotation $d\Theta_s/dE$, and
the dependence of the luminosity per electron on beam energy $d\lum(E)/dE$.

During the 1994/95 run, a number of measures in the operation of the SLC and in monitoring procedures significantly reduced the size of this {\it chromaticity}
correction and its associated error.
The fractional RMS beam energy spread was reduced to approximately 0.12\% (0.20\% in 1993) and non-Gaussian tails in the beam energy distribution were reduced to a negligible level\cite{decker}.  Optimization of the SLC arc spin transport system reduced the measured energy dependence of the spin rotation in the arc to $d\Theta_s/dE=1.4$~rad/GeV (2.5~rad/GeV in 1993).
Finally, $d\lum(E)/dE$ was reduced by improvements in the SLC
final focus optics \cite{FFupgrade}.  Constraints on $d\lum(E)/dE$
were made directly from
our data via a determination of the Z production rate as a function
of beam energy, with consistent results obtained from the observed
energy dependence of the beam size and from simulations of the
final focus optics \cite{FFupgrade}.
We then determine a contribution to $\xi$ of
$+0.0020\pm0.0014$ due to the chromaticity effect,
which is smaller by a factor of
eight than it was in 1993.
An effect of similar magnitude
arises due to the small
precession of the electron spin in the final focusing
elements between the SLC IP and the polarimeter.  This effect contributes $-0.0011\pm0.0001$ to $\xi$.
The depolarization of the electron beam by the \ee\ collision process is expected to be negligible \cite{chenyok}.  The contribution of depolarization to $\xi$ is determined to be 0.000$\pm$0.001 by comparing polarimeter data
taken with and without beams in collision.
Combining the three effects described above, the overall correction
factor is determined to be $\xi = 0.0009\pm0.0017$.

The $\ee$ collisions are measured by the SLD detector
which has been described elsewhere \cite{sld}.
The trigger relies
on a combination of calorimeter and tracking information;
the event selection is based on the
liquid argon calorimeter (LAC) \cite{lac} and the central drift
chamber tracker (CDC) \cite{cdc}.
For each event candidate,
energy clusters are reconstructed in the LAC.  Selected
events are required to contain at least 22~GeV of energy observed in the
clusters and to manifest a normalized energy
imbalance of less than 0.6 \cite{eimb}.  The left-right asymmetry associated
with final state $\ee$ events is expected to be diluted by the t-channel
photon exchange subprocess.  Therefore, we exclude $\ee$ final states
by requiring that each event candidate contain at least 4
selected CDC tracks,
with at least 2 tracks in each hemisphere defined with respect to
the beam axis, or at least 4 tracks in either hemisphere
(this track topology requirement excludes Bhabha events which contain
a reconstructed gamma conversion).
The selected CDC tracks are required to extrapolate to within 5 cm
radially and 10 cm along the beam direction of the IP, to have
a minimum momentum transverse to the beam direction of 100 MeV/c,
and to form a minimum angle of 30 degrees with the beam direction.

We estimate that the combined efficiency of the trigger
and selection criteria is (89$\pm$1)\% for
hadronic $\zo$ decays.  Tau pairs constitute (0.3$\pm$0.1)\% of the sample.
Because muon pair events deposit little energy in the calorimeter,
they are not included in the sample.
The residual background
in the sample is due primarily to
$\ee$ final state events.
We use our data and a Monte Carlo simulation to
estimate this background fraction
to be ($0.08\pm 0.08)\%$.  The background fraction
due to cosmic rays, two-photon events and beam related processes
is estimated to be (0.03$\pm$0.03)\%.

A total of 93,644 $\zo$ events
satisfy the selection criteria.
We find that 52,179 ($N_L$) of the events
were produced with the left-handed
electron beam and 41,465 ($N_R$)
were produced with the right-handed beam.
The measured left-right cross section asymmetry for $\zo$ production
is \cite{helicity}
\begin{eqnarray*}
\ameas\equiv(N_L-N_R)/(N_L+N_R)=0.11441\pm0.00325.
\end{eqnarray*}
We have verified that the measured asymmetry $\ameas$
does not vary significantly as more restrictive criteria (calorimetric
and tracking-based) are applied to the sample and that $\ameas$ is
uniform when binned by the azimuth and polar angle of the thrust axis.

The measured asymmetry $\ameas$ is related to $\alr$ by the following
expression which incorporates a number of small correction terms in
lowest-order approximation,
\begin{eqnarray}
\alr & = & \frac{\ameas}{\apolel}+\frac{1}
{\apolel}\biggl[f_b(\ameas-\aback)-\alum+\ameas^2\apol \nonumber \\
&  & -E_{cm}\frac{\sigma^\prime(E_{cm})}{\sigma(E_{cm})}\aengy
-\aeff + \apolel\polp \biggr], \label{eq:alrcor}
\end{eqnarray}
where $\apolel$ is the mean luminosity-weighted polarization for the
1994-5 run; $f_b$ is the background fraction;
$\sigma(E)$ is the unpolarized $\zo$ cross section at energy $E$;
$\sigma^\prime(E)$ is the derivative of the cross section with
respect to $E$;
$\aback$, $\alum$, $\apol$, $\aengy$, and
$\aeff$ are the left-right asymmetries \cite{asymdef}
of the residual background,
the integrated luminosity, the beam polarization,
the center-of-mass energy, and
the product of detector acceptance and efficiency, respectively;
and $\polp$ is any longitudinal positron polarization which is assumed to
have constant helicity \cite{ppol}.

The luminosity-weighted
average polarization $\apolel$ is estimated from measurements of
$\pole$ made when $\zo$ events were recorded,
\begin{equation}
\apolel = (1+\xi)\cdot\frac{1}{N_Z}\sum_{i=1}^{N_Z}{\poll}_i
=(77.23\pm0.52)\%, \label{eq:poldef}
\end{equation}
where
$N_Z$ is the total number of $\zo$ events, and ${\poll}_i$ is the
polarization measurement associated in time with the $i^{th}$ event.
The error on $\apolel$ is dominated by the systematic
uncertainties on the polarization measurement.

The corrections defined in equation~(\ref{eq:alrcor}) are found to be small.
The correction for residual background contamination is moderated by
a non-zero left-right background asymmetry ($A_b=0.055\pm0.021$)
arising from $\ee$ final states which remain in the sample.
Residual electron current asymmetry ($\lsim10^{-3}$) from the SLC polarized source was reduced by twice reversing
a spin rotation solenoid at the entrance to the SLC
damping ring.  The net luminosity asymmetry is estimated from the measured asymmetry of the rate of radiative Bhabha
scattering events observed with a monitor located in the North
Final Focus region of the SLC
to be $\alum=(-1.9\pm0.3)\times10^{-4}$.
A less precise cross check is performed
by examining the left-right asymmetry of the sample of 246,845 small-angle
Bhabha scattering events detected by
the luminosity monitoring system (LUM) \cite{berridge}.
Since the theoretical left-right asymmetry for small-angle
Bhabha scattering is very small [${\cal O} (10^{-4})\pole$
within the LUM acceptance], the measured
asymmetry of ($-$18$\pm$20)$\times$10$^{-4}$ is a direct determination
of $\alum$ and is consistent with the more precisely determined one.
The polarization asymmetry is directly measured to be
$\apol=(+2.4\pm1.0)\times10^{-3}$.
The left-right beam energy asymmetry arises
from the small residual left-right
beam current asymmetry due to beam-loading of the accelerator and is
measured to be (+9.2$\pm$0.2)$\times$10$^{-7}$.  The coefficient of the energy
asymmetry in equation~(\ref{eq:alrcor}) is a very sensitive function of the center-of-mass energy and is found to be $0.0\pm2.5$ for
$E_{cm}=91.280\pm0.025$~GeV.
The SLD has a symmetric acceptance in polar angle \cite{accept} which implies
that the efficiency asymmetry $\aeff$ is negligible.
As was discussed in our previous publication \cite{alr93}, the positron polarization at the SLC IP is less than 1.5$\times$10$^{-5}$.
The corrections listed in equation~(\ref{eq:alrcor}) change
$\alr$ by ($+0.2\pm 0.06$)\% of the uncorrected value.

Using equation~(\ref{eq:alrcor}), we find the left-right asymmetry
to be
\begin{eqnarray*}
\alr(91.28~{\rm GeV}) =
0.1485\pm0.0042({\rm stat.})\pm0.0010({\rm syst.}).
\end{eqnarray*}

The various contributions to the systematic error are summarized in
Table~\ref{table1}.
Correcting this result to account for photon exchange and for
electroweak interference which arises from
the deviation of the effective \ee\ center-of-mass energy
from the $\zo$-pole energy (including the effect of initial-state
radiation), we find the pole asymmetry $\alro$ and the effective
weak mixing angle to be \cite{ewcorr}
\begin{eqnarray*}
\alro & = & 0.1512\pm0.0042({\rm stat.})\pm0.0011({\rm syst.}) \\
\swein & = & 0.23100\pm0.00054({\rm stat.})\pm0.00014({\rm syst.})
\end{eqnarray*}
where the systematic uncertainty includes the uncertainty on the electroweak interference correction (see Table~\ref{table1}) which arises from the $\pm$25~MeV uncertainty on center-of-mass energy scale.
Combining this value of $\swein$ with our previous measurements
 \cite{oldalr,alr93} we obtain the value,
\begin{eqnarray*}
\alro & = & 0.1543\pm0.0039 \\
\swein & = & 0.23060\pm0.00050.
\end{eqnarray*}
This $\swein$ determination is smaller by
2.5 standard deviations than the recent average of 23 measurements performed
by the LEP Collaborations \cite{lepew}.

We thank the personnel of the SLAC accelerator department
and the technical staffs of our collaborating institutions for their
outstanding efforts on our behalf.
This work was supported by the Department of Energy; the
National Science Foundation; the Istituto Nazionale di Fisica
Nucleare of Italy;
the Japan-US Cooperative Research Project on High Energy Physics;
and the Science and Engineering Research Council of the United Kingdom.


%
%


%
%
\begin{table}
\caption{Systematic uncertainties that affect the $\alr$ measurement.  The uncertainty on the electroweak interference correction is caused by the $\pm$25~MeV on the SLC energy scale.}
\label{table1}
\begin{tabular}{lccc}
Systematic Uncertainty & $\delta\pole/\pole$~(\%) & $\delta\alr/\alr$~(\%) & $\delta\alro/\alro$~(\%)\\
\hline
Laser Polarization & 0.20 & & \\
Detector Linearity & 0.50 & & \\
Analyzing Power Calibration & 0.29 & & \\
Electronic Noise & 0.20 & & \\ \hline
Total Polarimeter Uncertainty & 0.64 & 0.64 & \\
Chromaticity and I.P. Corrections ($\xi$) &     & 0.17 & \\
Corrections in Equation~(\ref{eq:alrcor}) &  & 0.06 & \\ \hline
$\alr$ Systematic Uncertainty &    & 0.67 & 0.67\\
Electroweak Interference Correction &    &  & 0.33\\ \hline
$\alro$ Systematic Uncertainty &    &  & 0.75\\
\end{tabular}
\end{table}


\begin{references}
\bibitem{alr93} K.~Abe \etal, {\it Phys. Rev. Lett.} {\bf 73}, 25 (1994)
[hep-ex/9404001].
\bibitem{swdef} We follow the convention used by the LEP Collaborations
in {\it Phys. Lett.} {\bf B276}, 247 (1992).
\bibitem{top} CDF Collaboration: F.~Abe, \etal, {\it Phys. Rev. Lett.} {\bf 74}, 2626 (1995) [hep-ex/9503002]; D0 Collaboration: S.~Abachi, \etal, {\it Phys. Rev. Lett.}
{\bf 74}, 2632 (1995) [hep-ex/9503003].
\bibitem{accept} The value of $\alr$ is unaffected by decay-mode-dependent
variations in detector acceptance and efficiency
provided that the efficiency for detecting a fermion at
some polar angle (with respect to the electron direction)
is equal to the efficiency for detecting
an antifermion at the same polar angle.
\bibitem{oldslc} M.~Woods, AIP Conference Proceedings 343, 230 (1995).
\bibitem{source} R.~Alley, \etal, {\it Nuc. Inst. Meth.} {\bf A365}, 1 (1995).
\bibitem{strlat} T.~Maruyama \etal, {\it Phys. Rev.} {\bf B46}, 4261 (1992).
\bibitem{rot} T.~Limberg, P.~Emma, and R.~Rossmanith, SLAC-PUB-6210,
May 1993.
\bibitem{enspa} J.~Kent \etal, SLAC-PUB-4922, March 1989.
\bibitem{polarimeter} R. King, SLAC-Report-452; changes to the polarimeter for the 1994-95
SLD run are not described in this report and include a higher repetition rate
Nd:YAG laser, improved laser polarization diagnostics, and the addition of
a quadrupole magnet to the Compton spectrometer magnets.
\bibitem{comref} See S.B.~Gunst and L.A.~Page, {\it Phys. Rev.} {\bf 92},
970 (1953).
\bibitem{levchuk} M.~Swartz \etal, {\it Nucl. Instr. Meth.} {\bf A363},
526 (1995) [hep-ex/9412006].
\bibitem{decker} F.-J. Decker, R. Holtzapple, and T. Raubenheimer,
Proceedings of the 17th International Linear Accelerator Conference,
Tsukuba, Japan (1994), p.~47.
\bibitem{FFupgrade} F.Zimmermann \etal, SLAC-PUB-95-6790, June 1995.          
\bibitem{chenyok} P.~Chen and K.~Yokoya, {\it Proceedings of the Eighth International Symposium on High-Energy Spin Physics}, Minneapolis, MN, 1988, pg.~938.
\bibitem{sld} The SLD Design Report, SLAC Report 273, 1984.
\bibitem{oldalr} K.~Abe \etal, {\it Phys. Rev. Lett.} {\bf 70}, 2515 (1993).  Details of the calorimetric event selection can be found in J. Yarmartino, SLAC REPORT 426, February 1994.
\bibitem{lac} D. Axen \etal, {\it Nucl. Instr. Meth.} {\bf A328}, 472 (1993).
\bibitem{cdc} M. Fero \etal, {\it Nucl. Instr. Meth.} {\bf A367}, 111 (1995).
\bibitem{eimb} The energy imbalance is defined as a normalized vector
sum of the energy clusters as follows,
$ E_{imb}=|\sum \vec E_{cluster}|/\sum |E_{cluster}|$.
\bibitem{helicity} The absolute sign of $\ameas$ is inferred from the sign of the
measured Compton scattering asymmetry, the
measured helicity of the polarimeter laser, and the theoretical sign
of the Compton scattering asymmetry.
\bibitem{asymdef} The left-right asymmetry for a quantity $Q$ is defined as
$A_Q\equiv(Q_L-Q_R)/(Q_L+Q_R)$ where the subscripts $L$,$R$ refer to
the left- and right-handed beams, respectively.
\bibitem{ppol} Since the colliding electron and positron bunches
are produced on different machine cycles and since the
electron helicity of each cycle is chosen randomly, any positron
helicity arising from the polarization of the production electrons
is uncorrelated with electron helicity at the IP.  The net positron
polarization from this process vanishes rigorously.  However,
positron polarization of constant helicity does affect the measurement.
\bibitem{berridge} S.C. Berridge \etal,
{\it IEEE Trans. Nucl. Sci.} {\bf NS-39}, 242 (1992).
\bibitem{ewcorr} The quantities $\alro$ and $\swein$ are related by
equation~(\ref{eq:alrswein}) and are completely equivalent.  The correction for electroweak interference and pure photon exchange, $\alro-\alr(91.280)$ is determined with the ZFITTER~4.9
program of D.~Bardin, \etal\ (CERN-TH.~6443/92, May~1992) and is found to be $0.00265\pm0.00049$.
\bibitem{lepew} A.~Blondel, {\it Proceedings of the XXVIII$^{th}$ International Conference on High Energy Physics}, 25-31 July 1996, Warsaw, Poland.

\end{references}
\end{document}